\documentstyle[aps,tighten,twocolumn,epsf]{revtex}
\begin{document}
\twocolumn
%
\wideabs{
\title{Aging in an infinite-range Hamiltonian
 system of coupled rotators}

\author{Marcelo A. Montemurro$^1$, Francisco A. Tamarit$^1$ and Celia Anteneodo$^2$}
\address{$^1$
         Facultad de Matem\'atica, Astronom\'\i a y F\'\i sica,
         Universidad Nacional de C\'ordoba, \\
         Ciudad Universitaria, 5000 C\'ordoba, Argentina \\
         {\rm e-mails: mmontemu@famaf.unc.edu.ar, tamarit@fis.uncor.edu } \\
         $^2$
         Centro Brasileiro de Pesquisas F\'{\i}sicas\\
         Rua Xavier Sigaud 150, 22290-180, Rio de Janeiro, Brasil \\
         {\rm e-mail: celia@cbpf.br} }

\date{\today}

\maketitle

\begin{abstract}
We analyze numerically the out-of-equilibrium relaxation dynamics
of a long-range Hamiltonian system of $N$ fully coupled rotators.
For a particular family of initial conditions, this system is known
to enter a particular regime in which the dynamic behavior
does not agree with thermodynamic predictions. 
Moreover, there is evidence that in the thermodynamic
limit, when $N\to \infty$ is taken prior to $t\to \infty$,  
the system will never attain true equilibrium. By analyzing the
scaling properties of the two-time autocorrelation function we
find that, in that regime, a very complex dynamics unfolds, in which 
{\em aging} phenomena appear. The scaling law strongly suggests that 
the system behaves in a complex way, 
relaxing towards equilibrium through intricate trajectories. The present
results are obtained for conservative dynamics, where there is no
thermal bath in contact with the system. 
This is
the first time that aging is observed in such Hamiltonian systems.

\end{abstract}

 }
\narrowtext

At the very foundations of statistical mechanics, there are still
some hypotheses whose validity rests merely on the extrapolation
of observational facts and that have to be justified {\it a
posteriori}. Among them, let us mention two assumptions that are
intimately related to the issue addressed in this paper. The first
one refers to the introduction of a probabilistic description of
the evolution of a physical system. The second is related to the
mechanical specifications that a system must fulfill so that the
results of statistical mechanics can be applied \cite{krylov}.
These two points are closely related to the fundamental problem of
establishing a connection between the dynamical behavior of a
system, described by the Hamiltonian $H$, and its thermodynamics.
In that sense, statistical mechanics requires the existence of
adequate conditions allowing to replace the dynamical temporal
predictions by a probabilistic ensemble calculation that yields
the correct equilibrium mean value of the relevant quantities.

A very fast relaxation and a high degree of chaos and mixing are
usually required in order to guarantee that a system orbit will
cover most of its phase space in a short time. These questions
have been extensively investigated for low dimensional systems,
whereas, for extended systems, with infinite degrees of freedom,
the matter is still far from settled \cite{carati}.

Due to the analytical and numerical efforts of many authors, it
is today a well established fact that even very simple models,
when analyzed in the thermodynamical limit $N\to \infty$, can
yield results that bring to surface central questions about the
foundations of statistical mechanics. In particular, in this 
work, we will concentrate on an infinite-range (mean-field)
Hamiltonian system  that, despite its simplicity, exhibits a very
peculiar dynamical behavior: depending on the initial preparation
of the system its evolution can get trapped into trajectories that
will prevent the system from attaining equilibrium in 
finite time when the $N\to \infty$ limit is taken before the  
$t\to \infty$ limit \cite{antoni,vito,qqs_l,qqs_t,qqs_all,dyn_qqs}.
Unlike most models that display complex macroscopic behavior, this
infinite-range model includes neither randomness nor frustration
in its microscopic interactions. Furthermore, on one hand it
can be exactly solved in the canonical ensemble, while on the other hand 
it can be efficiently integrated in the microcanonical ensemble.
In that sense, it is an excellent starting point for analyzing the
above mentioned basic questions.

The system consists of $N$ fully coupled rotators whose dynamics
is described by the following Hamiltonian:
\begin{equation}
{\cal H}=\frac{1}{2}\sum_i L_i^2+\frac{1}{2N} \sum_{i,j} \Bigl[
1-\cos(\theta_i-\theta_j) \Bigr] \equiv K+V.
\end{equation}
It is worth mentioning that this is a rescaled version of a
nonextensive infinite-range Hamiltonian \cite{anteneodo}.
However, both of them share the same dynamical behavior after
appropriate rescaling of the dynamic variables.

In Fig. \ref{caloric}, we display the plot, $T$ vs. $U$, where
$T=2\langle K \rangle /N$ is the temperature and $U=(K+V)/N$ is the
total energy per particle. The solid line corresponds to the
canonical calculations \cite{antoni}, which predict a second-order
transition at $U_c=0.75$. Above $U_c$  constant specific heat is
found and below $U_c$ the system orders in a clustered phase. The
symbols correspond to the numerical results obtained by
integrating the equations of motion for $N=1000,\, 5000$ 
rotators and until $t=1000$. The system is initially prepared in a
``water-bag'' configuration, that is,  all the angles are set to
zero while the momenta are randomly chosen from an uniform
distribution such that the system has total energy $NU$. 
By measuring the nonequilibrium temperature (or equivalently the 
magnetization) of a system started in these  out-of-equilibrium initial 
conditions, one observes that, for a range of energy values 
below the transition, the system enters in a {\em quasistationary}  regime
characterized by a mean kinetic energy that  
varies very slowly. Moreover, the value of this non-equilibrium
temperature remains different from that predicted by canonical calculations. 
Actually, standard equilibrium is attained only after a time which grows 
with the size of the system, hence
an infinite system will never reach true equilibrium \cite{qqs_all,dyn_qqs}. 
In the quasistationary regime preceding equilibrium, 
trajectories are non-ergodic and the dynamics is weakly chaotic with 
Lyapunov exponent vanishing in the thermodynamic limit \cite{qqs_l}.  

It is our objective here to show that the discrepancy between the results 
drawn from the dynamics and those derived from the canonical ensemble
is closely associated to the presence of strong 
long-term memory effects and slow relaxation dynamics, a
phenomenon usually  named {\em aging}. Aging is one of the most
striking features in the off-equilibrium dynamics of complex
systems. It refers to the presence of strong memory effects
spanning time lengths that in some cases exceed any available
observational time. Although aging has been seen in a wide variety
of contexts and systems \cite{bouchaud}, some of them, actually
very simple ones \cite{carati,ritort}, it is perhaps in the realm
of spin glass dynamics where a systematic study of these phenomena
has been carried out (see Ref., \cite{bouchaud} and references therein).
Systems that {\em age} can be classified into dynamical
universality classes according to the scaling properties of their
relaxation function. Moreover, these scaling properties contribute
to a quantitative description of complex phenomena, even in cases
where a general theory is lacking \cite{bouchaud,yoshino}.

Aging can be characterized by measuring the two-time
autocorrelation function along the system trajectories. If the
state of the system in phase space can be completely characterized
giving a state vector $\vec{x}$, then the two-time autocorrelation
function is defined as follows:
\begin{equation}
C(t+t_w, t_w)= \frac {\langle {\vec{x}(t+t_w)} \cdot
{\vec{x}(t_w)}\rangle - \langle {\vec{x}(t+t_w)} \rangle \cdot
\langle {\vec{x}(t_w)} \rangle}
 {\sigma_{t+t_w} \sigma_{t_w}},
\label{correlacion}
\end{equation}
where $\sigma_{t'}$ are standard deviations and the symbol
$\langle \cdots \rangle$ stands for average over several
realizations of the dynamics. In the case of a Hamiltonian system
with $N$ degrees of freedom, the state vector is decomposed in
coordinates and their conjugate momenta, therefore we establish
the following notation: $\vec{x} \equiv (\vec{\theta},\vec{L})$.

For systems that have attained ``true'' thermodynamical
equilibrium, only time differences make physical sense when
calculating relaxation quantities. In this case, it is expected
that on an average the system will show only very short memory of
past configurations. However, for systems exhibiting {\em aging},
a complex time dependence is observed in the behavior of $C(t+t_w,
t_w)$, indicating long-term memory effects. In such a case, even at
macroscopic time scales, the two-time autocorrelation function
shows an explicit dependence on both times ($t$ and $t_w$)
together with a slow relaxation regime.

In order to integrate the motion equations numerically, we employed
a fourth order symplectic method \cite{yoshida} with a fixed time
step selected so as to keep a constant value of the energy within
a relative error $\Delta E /E$ of order $10^{-4}$. All the
simulations were started from the water-bag initial conditions
explained above.

In Fig. \ref{ene069}, we present the results of the numerical
calculation of the two-time autocorrelation function
(\ref{correlacion}) for $U=0.69$. This value of the energy
together with the water-bag initial conditions set the system into
a particular dynamical regime, in which ensemble discrepancy is
more pronounced when finite size results are extrapolated to the
thermodynamic limit. In the graph, features usual of aging
phenomena can be distinguished. For a given $t_w$ the system 
first enters a quasiequilibrium stage, 
in the sense that temporal translational invariance holds, 
with $ C(t+t_w,t_w) \approx$ 1,  
up to a time of order $t_w$. After that,  the system enters a
second relaxation characterized by a slow power law decay and a
strong dependence on both times. This phenomenology can be clearly
seen in the curves for the largest $t_w$'s.

In Fig. \ref{collapse}, we show the best data collapse for the
long-time behavior of the autocorrelation function, using  the
data of Fig. \ref{ene069} corresponding to the three largest
waiting times ($t_w=2048$, $8192$, and $32768$). The resulting
scaling law clearly indicates that for the whole range
of values of $t/t_w$ considered:
\begin{equation}
C(t+t_w,t_w) = f \Bigl( \frac{t}{t_w^{\beta}} \Bigr)  \, ,
\label{scaling}
\end{equation}
where $f(t/t_w^\beta)\sim (t/t_w^\beta)^{-\lambda}$. 
It is worth mentioning that this scaling 
is the same as observed experimentally 
in spin glass systems \cite{Hammann}.
The values obtained for the scaling parameters are 
$\beta\approx 0.90$ and $\lambda \approx 0.74$.  
In the inset, we exhibit an
alternative representation of the data which yields a linear plot.
It corresponds to $\ln_q[C(t+t_w, t_w)]$  vs. $t/t_w^\beta$, 
where the function $\ln_q(x)$, named $q$-logarithm, is defined as
follows \cite{tsallis_lnq}:
\begin{equation}
\ln_q(x)=\frac{x^{1-q}-1}{1-q}\quad .
\label{ln_q}
\end{equation}
In this expression, $q=1+1/\lambda$, which for the data in Fig.
\ref{ene069}, yields $q \approx 2.35$. Therefore, we can obtain a
complete functional form of the autocorrelation function valid
over the whole range of the scaling variable $t/t^\beta_w$ just by
identifying the function $f(x)$ in Eq. (\ref{scaling}) with the
inverse of $\ln_q(x)$, that is, the $q$-exponential 
\cite{tsallis_lnq}:
\begin{equation}
f(x) \equiv e_{q}^{-x} = \Bigl[ 1 - (1-q)x \Bigr]^{\frac{1}{1-q}},
\end{equation}
which naturally arises within the nonextensive statistics introduced 
by Tsallis inspired by the probabilistic description of multi-fractal 
geometries  \cite{tsallis}. 
The same qualitative behavior has been observed for other systems
sizes, namely $N=500, 2000$.

This aging scenario contrasts with the time invariant behavior observed 
within the high energy phase, where no quasistationary regime is detected. 
In fact, let us discuss Fig. \ref{ene5} where we present the results of 
the calculation of the two-time autocorrelation function (\ref{correlacion}) 
for $U=5.0$, well above the second order phase transition (i.e., inside
the homogeneous phase), with water-bag initial conditions.
What we observe here is essentially that 
the autocorrelation function depends on the two times only 
through their difference, that is, $C(t+t_w,t_w)\approx C(t)$. 
Therefore, the presence of aging is
related to the existence of quasistationary states.
It is important to emphasize that,
although the dynamics presents temporal translational invariance 
in the high energy regime, the relaxation of the system is very slow.

There is nowadays growing evidence that aging is a very common
dynamical phenomenon, associated to a great variety of physical
systems. So far, there are two scenarios within which aging can
emerge. On the one hand, the onset of aging in many systems
derives from the presence of coarsening processes that give place
to critical slowing down of the dynamics. In this case the scaling
law of the two-time autocorrelation function is ruled by the
following expression
\begin{equation}
C(t+t_w,t_w) \sim f \bigl( L(t)/L(t_w) \bigr),
\end{equation}
where $L(t)$ is the mean linear size of the domains at time $t$ 
\cite{bray}. On the other hand, aging also appears as a
onsequence of weak ergodicity breaking \cite{bouchaud} and it is
related to the complex fractal structure of the region of phase
space that the system explores in time. This is the case, for
instance, in the Sherrington-Kirkpatrick (SK) model and other
spin glass models in which the complexity of the energy landscape
is associated to a certain degree of randomness and/or frustration
in the Hamiltonian.

What is particularly remarkable in this work  
is the appearance of a complex
aging behavior in a {\em mean-field} model lacking both
randomness and frustration. Since the model we are analyzing in
this paper is an infinite-range one, such as the SK model, a
coarsening scenario has to be ruled out from the outset. 
Furthermore, all our results are obtained in a
conservative system without any thermal bath in contact with
it.  

Moreover, it is worth noting that the scaling law found for the two-time
autocorrelation functions below the transition,  
where a quasistationary regime is detected, points to a scenario very
similar to that observed in spin glasses \cite{Hammann}. 
As occurs in spin glasses, there is weak breakdown of ergodicity  
which is consistent with
the observation of weakly chaotic orbits, i.e., with a vanishing
Lyapunov exponent in the thermodynamic limit \cite{qqs_l}.
Drawing the analogy with spin glasses even further, our results
seem to confirm that the system visits phase space confined inside
very intricate trajectories (presumably nonergodic). This
conjecture is also supported by features observed in
$\mu$ space \cite{qqs_t,qqs_all,dyn_qqs}. 
The fact that the relaxation of the two-time autocorrelation function
can be well fitted by a $q$-exponential decay over 
the whole range of $t/t_w$ deserves further investigation.  
Although a possible connection with 
nonextensive statistics \cite{tsallis} is still not clear, we believe that it 
would be interesting to examine this possibility.

In summary, in this paper we have characterized the slow relaxation 
dynamics of a long-range Hamiltonian system through  its aging dynamics.
Our   observation  of  the  existence  of  aging  in this 
Hamiltonian system  and its  characterization   by
scaling properties reminiscent of  spin glasses 
is a result that can contribute to establish 
a unified frame for the discussion of the 
out-of-equilibrium dynamics  of  systems  with  many  degrees  of  
freedom.

We thank Constantino Tsallis for very stimulating discussions. 
We also want to thank  Andrea Rapisarda, Vito Latora, Alessandro Pluchino
and Raul O. Vallejos for very useful comments.
This work was partially supported by grants from  CONICET
(Argentina), Agencia C\'ordoba Ciencia (Argentina), SEYCT/UNC
(C\'ordoba, Argentina), and FAPERJ (Rio de Janeiro, Brazil).

\begin{figure}
\caption{The full line corresponds to the canonical theoretical
caloric curve and the symbols correspond to numerical simulations
for systems of size $N=1000$ (circles) and $N=5000$ (triangles),
at $t=1000$, averaged over 50 realizations. Initial conditions are
``water-bag''.  } 
\label{caloric}
\end{figure}

\begin{figure}
\caption{ Two-time autocorrelation function $C(t+t_w,t_w)$ vs. $t$
for systems of size $N=1000$ and energy per particle $U=0.69$. The
data correspond to an average over  200 trajectories initialized in
water bag configurations. The waiting times are $t_w$=8, 32, 128,
512, 2048, 8192, and 32768.} \label{ene069}
\end{figure}

\begin{figure}
\caption{Data collapse for the long-time behavior of the
autocorrelation function $C(t+t_w, t_w)$. The data are the same
shown in Fig. 2 for the three largest $t_w$. The gray solid line 
corresponds to  $e_q(-0.2\,t/t_w^{0.9})$. 
Inset: $\ln_q$-linear
representation of the same data, with $q\approx  2.35$.}
\label{collapse}
\end{figure}

\begin{figure}
\caption{Two-time autocorrelation function $C(t+t_w,t_w)$ vs. $t$
for systems of $N=1000$ and $U=5.0$. The data correspond to an
average over  10 trajectories initialized in water bag
configurations. The waiting times are $t_w$=8, 32, 128, 512, 2048,
8192, and  32768. 
Inset: semi-log representation of the same data.
} \label{ene5}
\end{figure}

%

%
%

\end{document}